\documentclass[10pt]{article}

\usepackage[letterpaper,margin=1in]{geometry}
\usepackage[T1]{fontenc}
\usepackage{times}
\usepackage{amsmath}
\usepackage{amssymb}
\usepackage{booktabs}
\usepackage{graphicx}
\usepackage{siunitx}
\usepackage[hidelinks]{hyperref}
\usepackage[numbers,sort&compress]{natbib}
\usepackage{caption}
\graphicspath{{figures/}}

\title{\vspace{-2.5em}A Calibration Audit of a Gaia XP White-Dwarf\\
Main-Sequence Binary Catalog: How Much the BP-Band\\
Residual Contaminates at Its Bound and Above}

\author{
  Karan Akbari\,\thanks{\href{https://orcid.org/0009-0005-0550-4018}{ORCID 0009-0005-0550-4018}; \texttt{karanakbari14@gmail.com}}\\
  St.\ Xavier's College, Mumbai, India
}
\date{}

\begin{document}
\maketitle

\begin{abstract}
\noindent
A Gaussian-process classifier on Gaia DR3 XP spectra produced $\sim$30{,}000
white-dwarf main-sequence binary candidates \citep{li2025wdms}, with
probabilities but no likelihood. The corrected spectra agree with external
libraries to better than 2\% below 400~nm and 1\% redward \citep{huang2024xp},
and that residual sits where a hot white dwarf also adds flux. We asked whether
it contaminates the selection at its bound. It does, weakly. The
spurious rate runs about 1.7 times the baseline there. Put into the blue band,
2\% of total flux is a median 55\% local excess across our MS templates, 27
times that number read locally. We inject a blue taper clipped to that bound,
under the templates' published per-pixel errors, with a $\Delta\chi^2$ threshold
standing in for the classifier. The spurious rate is $0.09\pm0.02$ across 20
seeds on a 0.053 baseline, 19 of 20 above it. The unclipped 2\% taper, above the
redward bound, gives $0.11\pm0.03$. At that taper the amortized posterior's
90\% coverage falls to 0.85 from a clean 0.88 on the flat-model leg, with PIT
uniformity rejected at $p<10^{-5}$ in all four seeds. The selection fails in
bulk only above a 10--15\% local excess, and the rate reaches 0.96 near 50\%. A
model-comparison gate certifies binarity only above a white-dwarf flux share
near 0.05. At share 0.01, near the catalog's median of 0.0096, the
$\Delta\chi^2$ preference inverts at a 20\% local excess. Outside Huang's
color--magnitude box, 40\% of the
catalog, the post-correction residual is not measured.
\end{abstract}

\section{Introduction}
\label{sec:intro}

White-dwarf main-sequence (WD--MS) binaries are close pairs that have survived a
common-envelope phase or will enter one, and as a population they constrain that
phase and feed the cataclysmic-variable and type Ia supernova progenitor
channels. Large samples turn individual systems into population statements, so a
catalog of tens of thousands is useful only insofar as its membership can be
trusted. When the catalog is built by a classifier that returns a probability and no
likelihood, it cannot say how many of its candidates are real, and which ones.

Gaia DR3 delivered low-resolution BP/RP (XP) spectra for about 220 million
sources \citep{gaia2016mission, gaiadr3}, and white-dwarf main-sequence
binaries are one of the populations
those spectra can isolate, because a hot white dwarf adds a blue excess on
top of an otherwise main-sequence continuum. \citet{li2025wdms} trained a
Gaussian-process classifier on this signature and shipped a catalog of
$\sim$30{,}000 WD--MS binaries, each with a probability \texttt{prob\_binary}
and a single-versus-binary fit, cut at \texttt{prob\_binary}~$>0.8$. There is no
test of whether the model behind the classifier's training spectra describes the
source. The same group has
since compared single against binary fits on XP spectra directly
\citep{li2025binaries}, and other Gaia DR3 white-dwarf and WD--MS catalogs have
been assembled from SED fitting, astrometry, and ultraviolet excess
\citep{rebassa2025wdms, nayak2025wdms, perezcouto2025som, shahaf2024triage,
gentilefusillo2021wd, garciazamora2025rf}. This note audits one of these
catalogs.

A model systematic the classifier
was never trained on could manufacture candidates without lowering a per-object
goodness-of-fit. The XP BP band has such a systematic. Below 400~nm the
uncorrected XP spectra carry a color-dependent calibration error, with red
sources measured brighter than expected \citep{montegriffo2023xp}, and a separate
processing bias overestimates the mean $G_{\rm BP}$ flux of faint red sources in
the photometry \citep{riello2021gaia}. The corrected XP catalogs agree with external
spectral libraries to better than 2\% over 336--400~nm and better than 1\%
redward \citep{huang2024xp}, which bounds the residual there without measuring
it. Whatever residual survives
lives in the same band, and points the same way, as the hot-white-dwarf excess the
classifier is built to find, so a single star carrying it could imitate a genuine
WD--MS binary.

At 2\% blue and 1\% redward, the BP systematic raises the spurious rate from
0.053 to 0.09 under the published per-pixel errors. It takes a local excess in
the tens of percent before spurious candidates come out in numbers. A
fraction of a star's total flux dumped into the narrow blue band is a much larger
\emph{local} excess than the same fraction read per wavelength, and on a red MS
star the gap is about a factor of 27
(Section~\ref{sec:silent}). We then locate the local excess that does start to
manufacture candidates, and we characterize how the posterior fails once it gets
there.

This is the optical counterpart of a test we ran on X-ray spectra
\citep{akbari2026xray}, where a 3\% detector gain shift slips past every
per-spectrum trust check and the evidence check alike, because the flow absorbs
the energy rescaling into the continuum parameters. That audit ended on a blind
spot. Here the residual is measurable in the selection at its bound,
and bulk manufacture needs local excesses in the 10--15\% range.
The injection and the model-comparison gate run on real Gaia
XP spectra, while the calibration test in Section~\ref{sec:calibration} uses an
analog neural posterior built for it, not the catalog's own classifier.

We forward-model the BP-band injection into real Gaia XP single-star spectra and
measure the spurious rate against the local excess. We test the catalog's
reliability against SDSS and LAMOST spectroscopy and GALEX ultraviolet
photometry. And we train an amortized neural posterior on the same forward model,
then run simulation-based calibration across injection
amplitudes. Section~\ref{sec:data} describes the catalog and the
spectra, Sections~\ref{sec:silent}--\ref{sec:calibration} give the three audits,
Section~\ref{sec:discussion} draws them together,
and Sections~\ref{sec:limitations}--\ref{sec:conclusion} set their scope.

\section{Data}
\label{sec:data}

The shipped catalog has 30{,}131 rows, all with \texttt{prob\_binary}~$>0.8$. This
is the post-cut FaintQC sample. The pre-cut pool of 452{,}433 candidates
\citep[their Section~4.3]{li2025wdms} is not distributed, so every catalog-level
statement here is certified within the 30{,}131 rows and says nothing about the
population the cut was drawn from. The MS--MS contamination flag \texttt{flag\_MSMS} is set for
7{,}621 rows (25.3\%), the authors' own model-based suspicion that a system is
two main-sequence stars. A majority, 56.1\%, have
\texttt{chi2\_diff\_renorm}~$<0$, meaning the single-star fit is at least as
good as the binary fit; consistent with the classifier's design, which targets
MS-dominated systems where the binary model does not improve $\chi^2$. The
binary-fit reduced $\chi^2$ has median 1.02, so under Li et al.'s error model the
fits are statistically reasonable where they are reported.

For the injection audit we built a data-driven forward model from real Gaia
DR3 \texttt{XP\_SAMPLED} spectra \citep{deangeli2023xp, montegriffo2023xp}:
120 single main-sequence stars and 120
single white dwarfs pulled with astroquery (with
\texttt{has\_xp\_sampled}~$=$~\texttt{true}), fit on a 50+50 subsample. The
MS side is an archive query on $BP-RP$ between 1.8 and 3.2, parallax above
10~mas, and \texttt{parallax\_over\_error}~$>50$, taking the 120 with the highest
\texttt{parallax\_over\_error}, which realizes 1.82--2.79 in
color and puts all 120 inside 100~pc. So the library sits nearby and red, with
100 of the 120 at $BP-RP\ge2$, outside the color box
\citet{huang2024xp} state their correction for. The
model space is the template libraries
themselves, following the classifier's own construction, and the injected
BP-band excess is the only misspecification we introduce.

For the reliability audit we cross-matched the catalog to SDSS
\citep{rebassa2016sdsswdms} and LAMOST \citep{ren2018lamostwdms} spectroscopic
WD--MS samples within $1.5''$. The SDSS side is the VizieR table
J/MNRAS/458/3808, which carries the 978 DR9--12 identifications of that paper
out of the 3{,}294 spectroscopic SDSS WD--MS binaries it reports in total.
Together they give 159 unique spectroscopically confirmed
WD--MS systems inside the catalog. We also matched to GALEX
\citep{martin2005galex, bianchi2017guvcat} through the CDS X-Match, which
covers 3{,}683 rows
(12.2\%) and supplies an ultraviolet detection independent of the optical
score.

\section{The injection and the amplitude it takes}
\label{sec:silent}

We inject the BP-band blue excess into single main-sequence spectra and refit
them with the full template library, varying the local amplitude of the excess.
To control the dominant real-data confound, that a sparse fit library cannot span
the variety of real single-MS spectra and would manufacture false positives on
its own, we fit each truth spectrum with its generating template excluded
(leave-one-out). The real MS templates are not perfectly clean on this axis:
their median absolute projection onto the injection shape is $0.13$--$0.16\%$ of
total flux across the runs here, about 4\% read as a local excess on the
392--420~nm convention of Appendix~\ref{app:severity}. That is twice the unclipped
2\% taper on the same convention, and about four times the clipped profile, which
sits at 1\% over most of that band.
That projection sits in both the clean
baseline and the injected copies, and the threshold behind the baseline
rate is the 95th percentile of the same clean singles, so it absorbs the
projection by construction and the quoted rates measure the injected excess on
top of it. The other direction is not controlled: a leave-one-out library that
already carries blue structure can absorb part of the injected excess as well,
which would push the quoted rates low, and we have not measured how far.

\citet{huang2024xp} report their corrected-XP consistency as better than 2\% in
336--400~nm and better than 1\% at redder wavelengths, which is a \emph{local},
per-wavelength bound, so we inject a local
fractional excess held under it. The distinction matters, because injecting 2\% of a
star's total flux into the blue band is a different quantity. The BP profile is
narrow and the blue flux of a red MS star is a small part of its total, so 2\% of
the total, deposited in the blue, is a median 55\% local excess across our
templates (Appendix~\ref{app:severity}), about 27 times a 2\% local residual.
Scaling to total flux would inject a $\sim$55\% local
excess while the label on it still reads 2\%.

Huang's 2\% and 1\% are consistency figures, the agreement between the corrected XP
spectra and the MILES and LEMONY libraries. Neither is a measured residual
amplitude. We inject a coherent one-signed excess held under those
figures, applied to every star. The
half-cosine taper peaks at 392~nm and decays to zero at 680~nm, so an amplitude
set from the 2\% blueward half alone carries more than 1\% out to 532~nm, above
the redward half of the same bound and by up to a factor of two over
402--532~nm. We therefore clip the profile to the bound itself, 2\% at
392~nm and 1\% wherever the taper would exceed it, and take that as the
Huang-consistent injection. A shape that held the 1\% cap past 680~nm and out to
992~nm would be a bigger injection than this, and none of our runs do that. On the
61-pixel 10~nm grid only 392~nm falls inside their 336--400~nm range, so in
substance the clipped result tests the 1\% redward cap over 402--532~nm
(Appendix~\ref{app:inject}).

Under the published per-pixel flux errors of the same
templates the clipped profile gives a spurious rate of $0.09\pm0.02$ over 20
seeds against a 0.053
baseline, above baseline in 19 of the 20. The $\pm0.02$ is the spread across
seeds, not an error on the mean; taken against the fixed baseline the excess
is $0.036\pm0.006$ ($t = 6.5$ on 20 seeds), so it is not a scatter fluctuation.
But that $\pm0.006$ is a seed-level standard error. Every seed redraws its fit
library from the same fixed 120-template parent, so the spread it is computed from
carries no template-parent sampling variance and understates the uncertainty at
the template level. The count, 19 of 20 seeds above baseline, escapes the size
of that scatter, though not the one parent behind it.
The unclipped 2\% taper
gives $0.11\pm0.03$, which brackets this taper shape from above but is not the
Huang-consistent number, and the largest unclipped amplitude of the same taper that
stays under both halves of the bound at every pixel is 1\%, which gives
$0.077\pm0.019$. A flat noise model
at SNR 30, on the unclipped 2\% taper, gives $0.08\pm0.01$. Those two runs share
their seeds and their RNG stream, so that comparison is paired as well: the mean
difference is $+0.032$ with a standard error of 0.007, positive in 15 of 20
seeds. The baseline is a construction constant. It is
0.053, 8 of 150, in every seed of every run here, because the threshold is the
95th percentile of the same clean singles that are then scored.

The 2\% is the summary figure, and \citet{huang2024xp} report discrepancies
reaching about 3\% in the blue around 400~nm for both libraries (their
Section~5.2); the raised blue-edge variant of the profile covers that case.
Raising the clipped profile's blue edge from 2\% to 5\% lifts the 392~nm pixel to
5\%, holds the 1\% cap out to 592~nm instead of 532~nm, and leaves the profile
above the 2\% version as far as 672~nm. The rate does not move: 0.086 against
0.089, the same within the seed scatter, while the unclipped taper goes from 0.110
to 0.200 over the same step. The extra red flux the 5\% version adds buys
nothing. That comparison cannot
single out the capped 402--532~nm shelf, since both clipped profiles
carry it unchanged, and nothing here decomposes the $\chi^2$ by wavelength. The
published per-pixel errors are why the redward part of the taper matters at all.
They sit below the flat model's single $\sigma$ over nearly the whole
grid, and the gap widens toward the red: the median fractional error across the
MS templates runs 16\% at 392~nm against 0.5\% at 792~nm, while the flat model has
one $\sigma$ per object. So on those errors the same fractional excess is worth
more $\chi^2$ past 400~nm than at the blue edge, by the square of that error
ratio. The 1\% cap on that shelf is softer than it
looks. Huang normalize the MILES spectra and their XP
counterparts by the mean flux over 442--458~nm before comparing (their
Section~5.2), which sits inside the shelf, so a broad one-signed offset across
that region is partly divided out of the comparison that sets the 1\%. The true
redward residual there could be larger than the bound reads. That is a reason to
keep the conditional caution below.

\begin{figure}[tbp]
  \centering
  \includegraphics[width=0.9\linewidth]{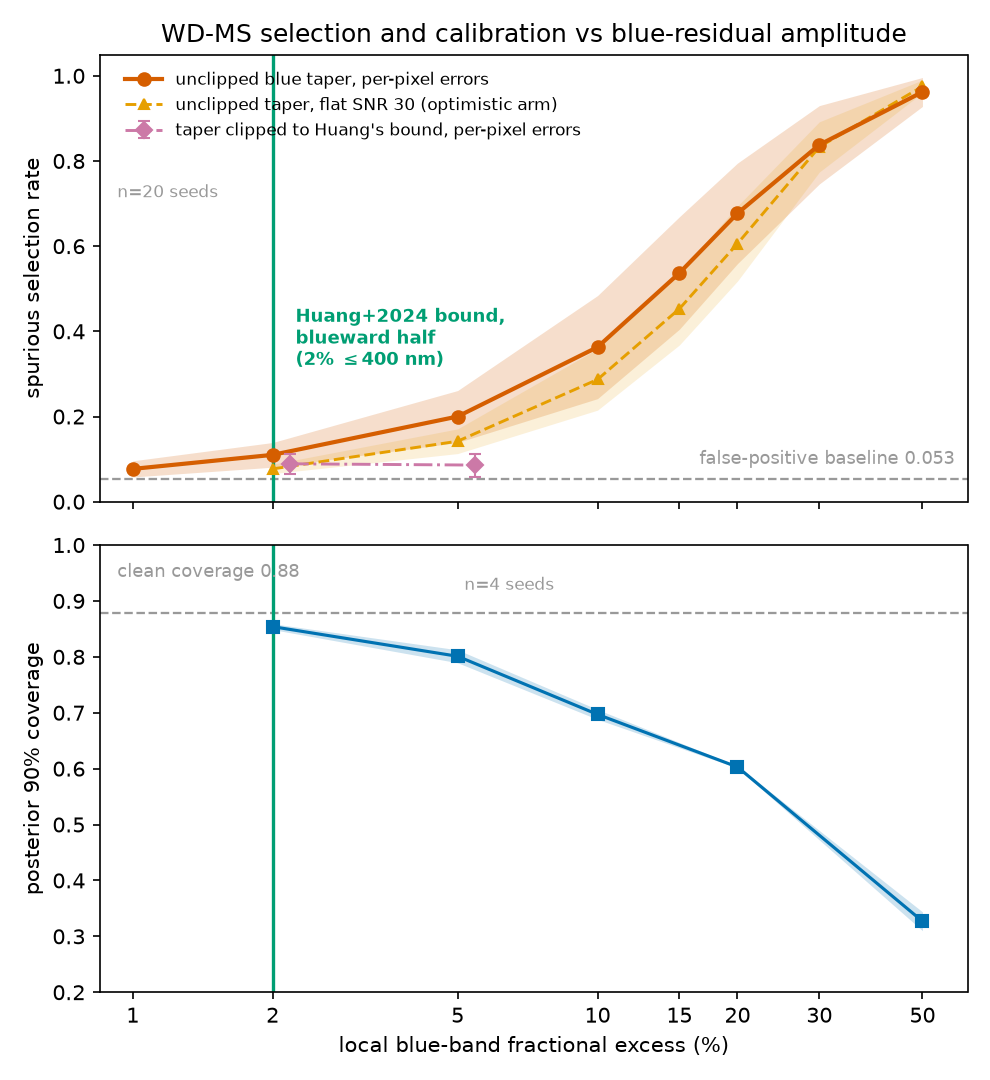}
  \caption{Spurious rate and posterior coverage against the local blue excess.
  \textit{Top:} the spurious
  WD--MS rate from injecting a blue excess into real single-MS Gaia XP spectra and
  refitting under leave-one-out, against the \emph{local} fractional
  excess in the blue band, over 20 seeds. Solid vermilion is the half-cosine taper
  of Appendix~\ref{app:inject} under the templates' published Gaia DR3 XP
  per-pixel flux errors, with its 1\% point from a separate run at the
  same seeds and per-seed thresholds; dashed orange is the same injection under a
  flat SNR-30 noise model; the dash-dot purple points are that taper clipped to
  Huang's two-part bound, 2\% at and below 400~nm and 1\% redward, again under the
  per-pixel errors, and they carry a small horizontal offset so they clear the
  curves. Only the left purple point is inside the whole bound: the right one is
  the same clipped profile at a 5\% blue edge, which is above the blueward half
  and clipped redward only. The green line marks the blueward half of the bound,
  where the taper
  peaks. There the clipped profile gives 0.09 and the unclipped taper 0.11, on the
  0.053 baseline (grey dashed). The unclipped per-pixel
  curve crosses 0.5 between a 10\% and a 15\% local excess and
  reaches 0.96 near 50\%. \textit{Bottom:} the amortized posterior's mean 90\%
  coverage over the same amplitude range (four seeds, flat SNR-30 noise model,
  measured at 2, 5, 10, 20 and 50\% local; the per-pixel-error reruns cover the
  selection leg only, so the coverage points are the unclipped taper throughout),
  from 0.85 at the unclipped 2\% taper
  (near the clean 0.88, grey dashed) down to 0.33 at 50\%. Shaded bands in both
  panels are $\pm1$ s.d.\ over seeds, as are the error bars on the two clipped
  points, and the amplitude axis is logarithmic.
  Both legs degrade over the same
  amplitude range, though they are matched in amplitude and their noise chains
  differ (Appendix~\ref{app:pca}).}
  \label{fig:turnon}
\end{figure}

The clean
threshold, set on those same published errors, rises with the extra $\chi^2$ they
buy (per-seed thresholds
0.29--0.55), so the net rate moves from 0.08 to 0.11 on the unclipped taper,
nothing like the raw $\chi^2$ gain. For that unclipped taper the rate rises with
the local excess, to 0.20 at 5\%, 0.36 at 10\%, 0.54 at 15\%, 0.68 at 20\%,
0.84 at 30\%, and 0.96 near 50\% (Figure~\ref{fig:turnon}; the flat-model curve
tracks it from below and the two meet at large amplitude). So the selection
manufactures candidates in bulk only once the local excess reaches the 10--15\%
range. Li et al.\ target the related $G_{\rm BP}$ photometric bias with a
$B<18$ selection, where $B$ is the same $G_{\rm BP}$ band, and their Section~2.2
prints the cut inside the parent query. But 24.5\% of the released rows sit at
$B>18$, and the catalog reaches $B\approx19.5$. We cannot tell from the release
which step relaxed it.
The 0.96 rate belongs to a regime far above the corrected residual. At the clipped
amplitude the elevation is about 1.7 times the baseline, 0.09 against 0.053.

Huang state their correction for
$-0.5<BP-RP<2$, $3<G<17.5$, and $E(B-V)<0.8$. On the $G$ axis the catalog is
nearly covered, with 6.5\% outside, but 38\% of it sits at $BP-RP\ge2$ and 40\%
falls outside the joint color--magnitude box, and color is the axis the blue
excess tracks. The $E(B-V)$ axis is not applied, because the catalog carries band
extinctions and no $E(B-V)$ column. It would not move the fraction anyway: the
largest $a_{\rm BP} - a_{\rm RP}$ in the 30{,}131 rows is 0.28~mag, which stays
under Huang's $E(B-V)<0.8$ for any reddening law with $R_{\rm BP} - R_{\rm RP}$
above 0.35, so applying that axis would remove no rows. For
that 40\% the post-correction residual is not measured. If it reaches the
10--15\% local range there, the
turn-on in Figure~\ref{fig:turnon} says the selection would start to fail. We
cannot rule that out, and we cannot show it either, so we state a
conditional caution for the part of the catalog outside the box, not a
contamination claim on it.

Faintness may be a second axis, though we have nothing that bounds it. Huang quote no
BP limit; their 17.5 is on $G$, and only 6.5\% of the catalog is outside on that
axis. Read across to $BP$ the same number would put 48.6\% outside, and most of
those rows are the red ones already counted: 53.7\% of them fail the color cut on
their own. The color cut leaves 20.3\% of the catalog, inside the joint
box but at $BP>17.5$, falling to 5.0\% at $BP>18$, which is Li et al.'s own
limit. If the blue-end residual grows as BP signal-to-noise falls, that is the
part it would reach, and we have no measurement of it either way.

Where it does manufacture spurious, the per-object statistics behave in ways a
downstream cut has to live with. First, the per-object $\Delta\chi^2$
is anti-informative for flagging
the spurious (AUC $\approx 0.0$; 0.002 across 20 seeds, at a 20\% local excess): the spurious singles keep
a lower $\Delta\chi^2$ than genuine binaries, so a $\Delta\chi^2$ cut ranks them
below the genuine binaries. Read the other way it separates them at 0.998 in the
same band, so the information is there and only the sign is wrong. The catalog
ships a related column, \texttt{chi2\_diff\_renorm}, which renormalizes the same
comparison differently and is not the statistic we score on
(Appendix~\ref{app:gate}), and Section~\ref{sec:reliability} finds the
fit does not drive the classification for the 56\% majority anyway, so how much of
this the released selection inherits is not something we can read off.
A plain binary-fit goodness-of-fit does regain power as
the excess grows (genuine-versus-spurious reduced-$\chi^2$ AUC $\approx 0.79$ at a
20\% local excess, $0.98$ at 50\%), which is why a bright-source magnitude cut
removes a systematic this large, but the classifier does not select on absolute
binary-fit $\chi^2$. Second, a model-comparison gate keeps power for the bright
synthesized binaries, but its power is set by the WD flux share. A BIC comparison
of the full binary model against a single star plus one free nuisance
flux-correction term separates genuine from spurious at AUC $\approx 0.93$ at a
20\% local excess. Those binaries are drawn across the whole bright band, WD flux
shares 0.048--0.231, the synthetic binaries' 0.05--0.30 flux-ratio band in share
terms.
Every AUC and spurious-flag rate in this section scores all 120 genuine binaries
against only the injected singles that passed the $\Delta\chi^2$ cut, the
would-be candidates, and the genuine side never goes through that cut. On the
$\Delta\chi^2$ axis the cut sits on the ranking statistic itself, so the
spurious class is truncated from below on its own score. A committed rerun
scores every injected single instead, on the same seeds and the same draws, and
its unconditioned numbers are printed beside the conditioned ones below. The
bright-band and basis-variation AUCs come from separate records that share the
construction and carry no unconditioned twin, though the rerun's own bright-band
arm moves 0.93 to 0.94 under the same switch.
The power comes from the comparison structure, so as a ranking statistic the gate
does not need to know the blue shape. Swapping the production blue ramp for a
flat, a red, or a random per-pixel nuisance template scores at least as well. In
the basis-variation run the gate AUC is 0.925 for the production ramp,
against 0.963 flat, 0.996 red, and 0.996 random, on the same 20 seeds at a 20\%
local excess. The separation survives the swap because a genuine binary needs a
white-dwarf shape a single free term cannot reproduce. At the random-basis limit the gate is close to the
per-object $\Delta\chi^2$ read backwards, 0.996 against the 0.998 the sign-flipped
$\Delta\chi^2$ gives on the same seeds. The fixed BIC-neutral
operating point is basis-dependent even though the ranking is not: the flagged
fraction of spurious singles falls from 0.66 under the production ramp to 0.02
under the random basis, so a threshold has to be recalibrated per basis. A
nested-sampling version of the same comparison, on the blackbody proxy at large
amplitudes, gives the same sign (Appendix~\ref{app:evidence}). But the gate's
genuine side falls off with the WD share. Restricted to the single fixed bin at share 0.05,
the AUC is 0.70, and 0.76 scoring all the injected singles. At the catalog's
typical share (median 0.0096) it is close to chance either way, 0.43 at share 0.01
against 0.54 unconditioned, because
the median gate statistic for a genuine binary flips sign, from $-31$ in the bright
band to $+3$ at share 0.01. The binary model is no longer preferred for the
catalog's typical genuine binary at this SNR. A BIC-neutral gate then flags about
66\% of the spurious (0.77 on every injected single) but rejects 85\% of the genuine (pass rate 0.15 at share 0.01
and 0.39 on the catalog's own share mix, against 0.93 in the bright band), so this
is a completeness collapse, not a purity
leak. Both flag and pass rates run through the fixed threshold, so both carry the
basis dependence above; the bright-band values are measured under all four bases
and the faint-share ones only under the production ramp. The $\Delta\chi^2$ statistic inverts across the same axis. Anti-informative
at bright shares (AUC $\approx 0.0$), it reaches 0.90 at share 0.01, or 0.71
unconditioned: the injected
spurious carry a larger $\Delta\chi^2$ improvement than the catalog's typical
genuine binaries, so a $\Delta\chi^2$-ranked selection prefers the spurious to the
faint real binaries it is meant to find. Drawn from the catalog's own share
distribution instead of a fixed share, the same comparisons land at 0.61
($\Delta\chi^2$) and 0.58 (gate), or 0.51 and 0.66 unconditioned, so the
fixed-share bins bracket the catalog mix.
The sign flip is a property of the faint end, and the mixture sits between the
regimes. The plain reduced-$\chi^2$ goodness-of-fit is the one per-object check
whose partial power survives at every share
(AUC 0.70--0.78 across the share arms at a 20\% local excess, 0.74--0.81
unconditioned), and the classifier
does not select on it. This adds a second, independent obstacle to the
prior-driven majority of Section~\ref{sec:reliability}, inside the injection
framework: at the catalog's typical
WD share no per-object statistic certifies binarity at this SNR, and the only
retained power is the goodness-of-fit flagging the spurious, not any statistic
confirming a faint genuine binary. Both gate/AUC runs are committed, the
conditioned one and its unconditioned twin (20 seeds,
shares 0.005--0.05 plus the catalog distribution and the bright band, amplitudes
10--50\%). Every AUC quoted since the turn-on, the basis variation included, runs
under the flat SNR-30 noise model. The published per-pixel errors enter the
turn-on curve alone.

\section{Catalog reliability versus external ground truth}
\label{sec:reliability}

The part of the catalog that external data can corroborate is the
$\chi^2$-favoring minority. The 159 spectroscopically confirmed WD--MS systems
have a median renormalized $\Delta\chi^2$ of $+0.72$, with only 11\% below
zero, whereas the full catalog has median $-0.018$ and 56\% below zero. The
confirmed systems are the ones where the binary model already wins on $\chi^2$.
The classifier's distinctive feature is the rest: the MS-dominated majority
where $\Delta\chi^2<0$ and the prior, not the fit, drives the classification.
\citet{li2025wdms} themselves note that the classifier's boundaries are shaped
by the training-data priors, affecting both completeness and reliability (their
Section~4.4). That majority is externally unverified. The reach of the external data is the
reason: only 20 of the SDSS and 146 of the LAMOST spectroscopically confirmed
WD--MS systems fall inside the catalog, with the 201 matching LAMOST entries
collapsing onto 146 distinct sources (159 unique across both surveys, 7 confirmed
by both). The data do not show the majority is wrong. Enlarging
the external sample does not change this, and the channels carry different
meanings, so we keep them separate (Table~\ref{tab:external}). A match
to the Gaia EDR3 white-dwarf catalogue \citep{gentilefusillo2021wd} certifies
only a white-dwarf candidate near the cooling track at that position, and for a
single-white-dwarf catalogue that can mean a lone white dwarf, so the match reads
as a contamination signal. Of the 220 matches, 18 clear the $P_{\rm WD}>0.75$
guideline the catalogue itself gives. The astrometric orbits of
\citet{shahaf2023triage1} add no white-dwarf-classified companion here, with 0 of
32 matches classified as such, a 95\% Wilson upper bound of 0.11. Their own base
rate is 101 white-dwarf labels over 101{,}380 rows, 0.10\%, so 32 matches expect
0.032 of them and the zero carries no information. No
external catalog reaches even 1\% of the 30{,}131.

\begin{table}[t]
  \centering
  \caption{External catalogs that touch the WD--MS catalog, kept separate by
  what each certifies. Counts are distinct catalog sources matched within
  $1.5''$ or by Gaia source identifier; fractions are of the 30{,}131 catalog
  sources. The spectroscopic union removes 7 systems confirmed by both surveys.
  Gentile Fusillo et al.\ ship a candidate catalogue carrying a per-source
  probability $P_{\rm WD}$, and name $P_{\rm WD}>0.75$ as ``a generic
  guideline'' for a high-confidence sample; the 220 matches have median
  $P_{\rm WD}$ 0.10 and 18 of them clear that cut, which is the indented row.
  A low $P_{\rm WD}$ does not make those 220 junk: an unresolved WD--MS composite
  sits off the single-white-dwarf cooling track the probability is fitted
  against. The row gives a base rate. Read as a contamination count, it would be
  wrong.}
  \label{tab:external}
  \begin{tabular}{@{}llrr@{}}
    \toprule
    External set & Certifies & Matched & Frac. \\
    \midrule
    SDSS J/MNRAS/458/3808 \citep{rebassa2016sdsswdms}  & spec.\ WD--MS confirmed        & 20  & 0.07\% \\
    LAMOST \citep{ren2018lamostwdms}  & spec.\ WD--MS confirmed        & 146 & 0.48\% \\
    \quad spectroscopic union         & spec.\ WD--MS confirmed        & 159 & 0.53\% \\
    GF21 \citep{gentilefusillo2021wd} & WD candidate at that position  & 220 & 0.73\% \\
    \quad at $P_{\rm WD}>0.75$        & high-confidence WD candidate   & 18  & 0.06\% \\
    Shahaf \citep{shahaf2023triage1}  & astrometric binary present     & 32  & 0.11\% \\
    \quad WD-classified companion     & astrometric WD companion       & 0   & 0.00\% \\
    \bottomrule
  \end{tabular}
\end{table}

The MS--MS flag is a noisy contamination indicator. Confirmed WD--MS systems
carry \texttt{flag\_MSMS} at 40.9\% (Wilson interval 0.335--0.486), against 28.9\%
for a $G$- and color-matched control drawn from the non-confirmed catalog and
25.3\% for the full catalog. So the flag is elevated in the confirmed WD--MS
relative to a matched control, but the separation is weak, and part of the
elevation tracks the binary mass-ratio axis that the flag is itself partly
constructed from.

\begin{figure}[tbp]
  \centering
  \includegraphics[width=0.8\linewidth]{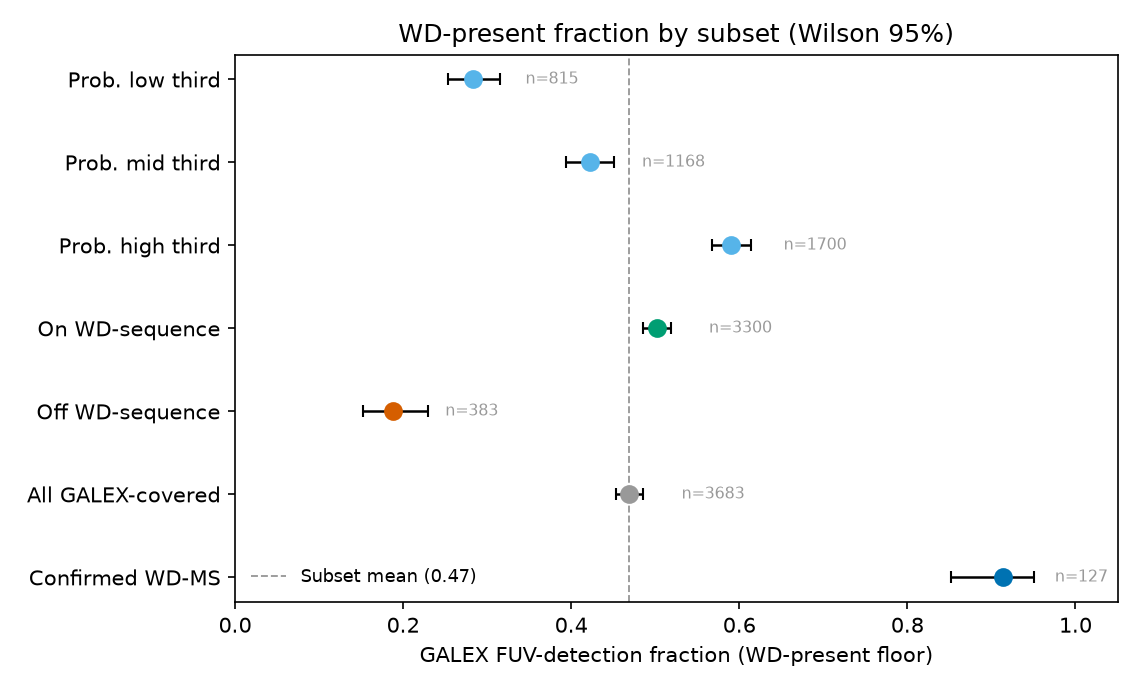}
  \caption{GALEX FUV-detection fraction, a floor on the fraction of each subset
  that hosts a hot white dwarf, with Wilson 95\% intervals and annotated subset
  sizes. FUV detection rises monotonically across \texttt{prob\_binary}
  terciles, from 0.28 in the lowest third to 0.59 in the highest, which is
  consistent with the score ranking true white-dwarf presence, while even the top
  tercile is only 59\%
  UV-confirmed. Terciles are cut on the full catalog's \texttt{prob\_binary}
  (boundaries 0.862 and 0.934). GALEX coverage rises across them (8.1\%, 11.6\%,
  16.9\% of each tercile), and the comparison is matched on nothing. So the trend
  is uncontrolled. Off-sequence sources (\texttt{flag\_wdmsfit\_in}~$=0$, about
  23\% of the catalog) are FUV-detected at 0.19 against 0.50 on-sequence, and the
  127 spectroscopically confirmed WD--MS systems show the white-dwarf UV
  signature at 0.91. The dashed line marks the 0.47 detection fraction over the
  full NUV-bright subset, a floor for that 12.2\% subset only.}
  \label{fig:reliability}
\end{figure}

Sources fit
off the white-dwarf cooling sequence (\texttt{flag\_wdmsfit\_in}~$=0$, about
22.8\% of the catalog) are ultraviolet-deficient: GALEX detects them in the FUV
only 19\% of the time, against 50\% for on-sequence sources. Off-sequence sources
are somewhat more distant (median 433 pc against 302 pc), so distance alone would
depress their FUV detection and could mimic the deficiency. Matching each
off-sequence source to a same-distance on-sequence source leaves the gap intact
and slightly wider (0.19 against 0.57, from 0.19 against 0.50 unmatched), so the
deficiency is not a distance artifact.
The 127 spectroscopically confirmed WD--MS systems with GALEX coverage show the
white-dwarf UV signature at 91\% (Figure~\ref{fig:reliability}). The UV label
enters neither the optical score nor the flag, so this indicator is
non-circular. The
MS--MS flag's two states do separate in the UV, in the direction opposite to what
the flag is supposed to mean (FUV detection 0.53 for flagged sources against 0.44
for unflagged), which is a further reason not to read it as a contamination
label. The indicator
detects the absence of a \emph{hot} white dwarf; a cool old white dwarf is also
FUV-faint, so it is a lower bound on white-dwarf presence, validated on the
GALEX-bright nearby subset. Li et al.\ already cross-matched this catalog to GALEX
and reported ultraviolet-excess fractions \citep{li2025wdms}, and the same
UV-excess logic drives other Gaia DR3 WD--MS searches \citep{nayak2025wdms}. The
off-sequence cut and the distance control behind it are new here.

\section{Calibration of the amortized posterior}
\label{sec:calibration}

The BP residual moves the selection modestly at its bound.
Amortized posteriors are starting to replace per-object fits, so we ran the same
injection against one and looked at how it fails at large amplitude. We trained an
amortized neural posterior
(a neural spline flow, sbi \citep{tejero-cantero2020sbi}) on a real-template PCA
forward model and ran simulation-based calibration \citep{talts2018sbc}. On clean
data, across four seeds the 90\% coverage runs
0.873--0.889 (mean 0.879). The correct null is not 0.90: the discrete rank estimator at 50
posterior draws places 45 of 51 PIT ranks inside $[0.05,0.95]$ under perfect
calibration, which gives 0.882, verified analytically and with a
conjugate-Gaussian self-test. So the clean flow sits on the target.

This leg injects the unclipped taper at its stated local amplitude
(Appendix~\ref{app:pca}), so at 2\% it runs above the redward half of Huang's
bound the way the selection leg's unclipped arm does, and the coverage damage it
reports is an upper bound on what a clipped profile would do. Coverage degrades
with amplitude over the same range as the selection. At the
unclipped 2\% taper the mean 90\% coverage is 0.85, a small drop from the clean
0.88 and a consistent one across the four seeds (0.848--0.861). It falls to 0.80 at 5\%, 0.70 at
10\%, 0.60 at 20\%, and 0.33 near 50\% (Figure~\ref{fig:turnon}, bottom). The drop
is small in coverage but not in the underlying ranks: the smallest
per-parameter Kolmogorov-Smirnov $p$-value against PIT uniformity is below
$10^{-5}$ in all four seeds at 2\%, while on clean data the same statistic is
consistent with uniformity (Appendix~\ref{app:sbc}). The posterior moves a little
in coverage at the amplitude that survives correction and fails only well above
it, the same way the selection does.

\begin{figure}[tbp]
  \centering
  \includegraphics[width=0.95\linewidth]{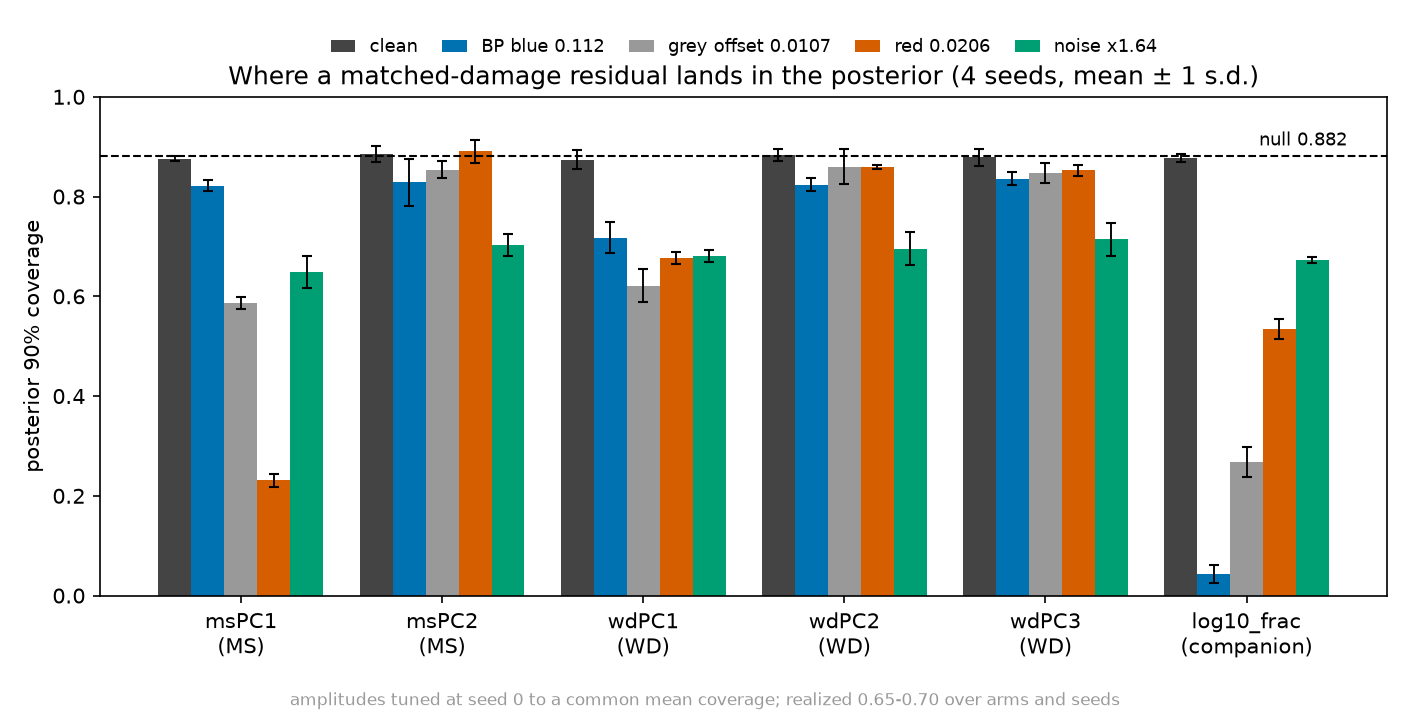}
  \caption{Where a matched-damage residual lands in the posterior. Bars are the
  mean 90\% coverage per parameter over four seeds ($\pm 1$ s.d.). The dashed line
  is the 0.882 discrete-rank null of Appendix~\ref{app:sbc}, and the measured
  clean coverage of 0.879 is marked in Figure~\ref{fig:turnon} instead. Each
  injection's amplitude is tuned once on the first seed to the same mean
  coverage damage and applied to all four (realized 0.648--0.704 across arms and
  seeds). The leftmost dark-grey bar of each group is the uninjected control,
  which runs 0.873--0.885 over the six parameters. The tuned amplitudes are a
  0.1124 local blue excess, a 0.0107 grey multiplicative offset, a 0.0206 red
  taper at its own peak, and a $\times1.638$ noise scale, so the blue arm is the
  largest perturbation of the four and 5.6 times the residual audited in
  Section~\ref{sec:silent}. The blue BP-shaped residual drives the companion
  fraction (\texttt{log10\_frac}) to 0.04 and leaves the leading main-sequence
  component at 0.82. Under the matched grey offset the two run 0.27 and 0.59. The
  matched red excess hits the main-sequence component instead (0.23) and leaves
  the companion fraction at 0.54. Matched extra noise degrades every parameter
  about evenly.
  The apparent main-sequence immunity under the blue arm is partly a basis effect:
  the leading MS component carries $<$6\% of its squared loading below 500~nm.}
  \label{fig:specificity}
\end{figure}

When the
posterior does miscalibrate, the damage concentrates on the companion
fraction. The concentration starts at small
amplitudes: the companion fraction's 90\% coverage is 0.79 at the unclipped 2\%
taper and 0.47 at 5\%, while the mean over parameters is still 0.85 and 0.80. At a 10\% local excess the companion
fraction's 90\% coverage drops to about 0.08 across four seeds while the leading
main-sequence component falls to 0.845, which is 0.037 below the 0.882 null. At
20\% the companion
fraction's coverage is 0.002 and the main-sequence component has itself fallen
to 0.79, so the damage still concentrates on the companion fraction while the
main sequence moves further off the null. A blue
excess is spectrally localized, so it aliases into the white-dwarf and companion
parameters that live in the blue, and it does this more than an equal-damage
perturbation of another kind. We matched a grey offset, a red-band excess, and
extra noise to the same mean coverage drop as the blue injection. The companion
fraction still collapses far more under the blue residual (coverage 0.04 across
four seeds) than under the matched grey offset (0.27), the matched red excess
(0.54), or matched noise (0.67; Figure~\ref{fig:specificity}). The four arms are matched
on coverage damage, so their perturbation sizes differ and
the blue one is the largest: it runs at a local excess of 0.1124, about 11\% at
the 392~nm peak, against a grey multiplicative offset of 0.0107, a red taper of
0.0206 at its own 992~nm peak, and a noise scale of $\times1.638$. That blue
amplitude is 5.6 times the residual under test and sits inside the 10--15\% band
where the selection itself starts to fail, so this experiment gives the shape of
a large residual's damage. At Huang's bound the posterior holds at least the 0.79
companion-fraction coverage quoted above, since that number comes from the
unclipped taper.
The tuning trace brackets each tuned amplitude, and the noise arm is the
steepest (mean coverage falls by about 0.44 per unit
$\ln a$, against 0.26 for the grey arm and 0.25 for the red), so its matched
number is the least stable. The blue arm is shallow, about 0.16. The noise arm
spreads its damage across every parameter about evenly. The posterior ends up
overconfident about the companion fraction. But it can still constrain it.
One caveat matters
here: the leading MS component carries almost no blue weight by
construction (0.055 of its squared loading below 500~nm), so its apparent immunity is
partly a property of the basis. The run gives the blue-specific, damage-matched
bias in the companion fraction, and says nothing about whether the main sequence
is physically untouched.

\section{Discussion}
\label{sec:discussion}

At the residual Huang's bound leaves, the injection raises the
spurious rate to $0.09\pm0.02$ against the 0.053 baseline under the published
per-pixel errors, and to 0.11 if the taper is held at 2\% past the point where
their bound drops to 1\% (0.08 under a flat SNR-30 model). On the
unclipped-taper flat-model leg the amortized
posterior holds a 90\% coverage of 0.85. The elevation is measurable, about 1.7
times the baseline, but bulk
manufacture needs a 10--15\% local excess. Huang's stated range does not cover
40\% of the catalog (Section~\ref{sec:silent}), so the audit gives only a
conditional caution there.

A catalog builder deciding whether a given correction is good enough needs a
number for how much residual it takes. Figure~\ref{fig:turnon} gives one, for this
selection. The posterior test (Figure~\ref{fig:specificity}) shows what such a
residual does when it miscalibrates an amortized posterior. It makes the companion
fraction confident and wrong
while the fit quality on the rest of the spectrum stays clean, and the
posterior-predictive diagnostic separates contaminated from clean inputs only
partially (AUC 0.79, one seed, on the blackbody pairs at an additive 5\%-of-total
injection, which on that population is about 23\% local). That failure mode, an untrained model
systematic breaking an amortized posterior's calibration while clean-data checks
pass, is the one studied in the simulation-based-inference robustness literature
\citep{hermans2022crisis, cannon2022misspec}, where SBC serves as the diagnostic
and robust or model-criticism methods as the fix \citep{ward2022rnpe,
anaumontel2025misspec}. Neural posterior estimation for white-dwarf spectroscopy
is already in use \citep{vincent2025dufourbergeron}, so the specific form the
failure takes for Gaia XP is worth having on record for the amortized successors
to catalogs like this one.

The reliability audit does not depend on the injection. The score appears to rank
true white-dwarf presence,
with FUV-confirmation rising across \texttt{prob\_binary} terciles from 28\% to
42\% to 59\%. That trend is uncontrolled, though: GALEX coverage rises across the
same terciles, from 8.1\% to 16.9\%, and unlike the off-sequence result below the
terciles are matched on nothing, so we cannot say how much of the gradient is the
score and how much is the subset it is measured on.

The probability does not certify purity in the formal sense. Conformal
prediction has been used in astronomy to calibrate gravitational-wave searches
\citep{ashton2024conformal}; the target here is catalog-level false-discovery
control. A conformal-FDR certificate \citep{jin2023conformal} using GALEX FUV
detection as an independent label, orthogonal to the optical score, finds that
the catalog's \texttt{prob\_binary} certifies no FDR-controlled selection at
$q\le0.2$. The 977 FUV-non-detected negatives in the calibration half floor the
conformal $p$-values at $1/978\approx10^{-3}$. That floor alone does not
explain the empty selection. A composite score (the probability and the $\chi^2$
preference rank-normalized, plus the on-sequence flag) selects 438 of the
2{,}706 GALEX-labeled objects in the same construction at $q=0.2$, under the
same floor, and over the full catalog it, too, selects none. That composite is an
existence check. It is not a performance number: one of its ingredients, the
on-sequence flag, was chosen because Section~\ref{sec:reliability} found it
predicts FUV detection on these same objects.
\texttt{prob\_binary}'s conformal $p$-values against the noisy negative pool
never clear the Benjamini--Hochberg line at any $q$ up to 0.2. The depth-limited FUV non-detection is too noisy a negative label to
support a tight null, since real white dwarfs below the GALEX depth contaminate
it. The certifiable statement, reading an FUV detection as a hot white dwarf,
is a binomial floor of about 47\% white-dwarf-present (95\% lower bound
45\%) within the 12.2\% NUV-bright GALEX subset. A tighter certificate needs a
cleaner negative, from deeper UV or spectroscopic non-WD--MS sources.

\section{Limitations}
\label{sec:limitations}

The injection is on real Gaia XP spectra under leave-one-out, with a
50+50 template fit library; the primary turn-on uses each template's published
per-pixel flux errors and a flat SNR-30 noise model is kept as the comparison
arm. Those flat errors are looser than the published per-pixel ones over almost
the whole grid. So the flat arm reads below the per-pixel one, 0.08 against 0.09
to 0.11 at 2\% local. It is an optimistic arm and brackets from one side only. The
coverage leg keeps that flat model, so it is optimistic in the same direction, by
an amount we have not measured. The per-pixel errors reach only the turn-on
curve, and the share-resolved gate and AUC runs keep that flat model too.
Each seed also draws its own 50-template fit library from one fixed
120-template parent, so the $\pm0.02$ and $\pm0.03$ of
Section~\ref{sec:silent} measure library draw and noise realization inside that
parent, with no population error bar behind them. They are also wide, 27\% and 26\% of their
own means. The $\pm0.006$ on the excess and the $t = 6.5$ it gives come from the
same scatter, so they understate the template-level uncertainty in the same way,
and the seed counts, 19 of 20 above baseline and 15 of 20 in the paired
comparison against the flat model, do not depend on how wide that scatter is.
They still rest on the one parent library. We do
not convert the injection into a catalog-wide spurious count, which would need the
residual-versus-magnitude relation for the released sample. That is the
main open piece. The conditional caution of Section~\ref{sec:silent} stands or
falls on whether the residual reaches the 10--15\% local range in the 40\% of the
catalog outside Huang's color--magnitude box, which their stated applicability
range does not cover. The selection statistic we
score on is a 95th-percentile $\Delta\chi^2$ threshold, a stand-in for Li's
Gaussian-process classifier, and the released classifier itself is never run here,
so the spurious rates describe that threshold. Our stand-in demands a positive
$\chi^2$ improvement above a high percentile while the released boundary admits
56\% of rows at a negative one. So the stand-in is the stricter of the two, and
0.09 does not bound what the real classifier would do at the same residual.
The 0.13\% template projection of Section~\ref{sec:silent} is not regenerable from
the current command line, because its record predates it. The GALEX cross-match
behind Section~\ref{sec:reliability} came from a live archive query with no cached
table behind it. And the two median per-pixel fractional errors quoted in
Section~\ref{sec:silent}, 16\% and 0.5\%, are printed by the committed loader at
run time and carried by no committed JSON, though a reader can regenerate those
from the committed script and the template manifest. Those three are the
exceptions to the committed-record standard the rest of the numbers meet.

The injection is also calibrated on templates in a regime where Huang's bound is
not measured, since 100 of the 120 MS templates sit outside their color box
(Section~\ref{sec:data}). The 55\% conversion and the rates that follow from it
are properties of those spectra.

Our single-MS truth sources are enforced by RUWE, \texttt{non\_single\_star},
and CMD position only, with no spectroscopy. Any WD--MS contamination among them
would dilute the injected effect, not create it, so the measured elevation is a
lower bound on the true one. It is
conservative for the claim that the elevation is real, and optimistic for the
claim that the catalog stays far from bulk manufacture.

The posterior-calibration test uses a real-template PCA forward model and an
analog neural posterior, so the
coverage numbers describe the amortized methods replacing per-object fits, not
Li et al.'s released catalog. The spectral-specificity arms are tuned to a common
coverage damage, so their perturbation magnitudes differ and the blue arm is the
largest of the four, and the result carries the basis caveat noted in
Section~\ref{sec:calibration}. The synthetic binaries also span WD flux fractions
of 0.05--0.30, while the catalog's own G-band WD flux fraction has a median near
0.0096, with about 64\% of sources below a 0.024 share, one of the bins of the
share-resolved run, so the coverage numbers are measured on binaries several times brighter
in WD fraction than the catalog median. The gate and AUC re-run at
catalog-representative fractions is in the committed record, and the
share-resolved result in Section~\ref{sec:silent} is measured there. Fractions in
that run are the WD share of total grid flux, which for the bluer WD sits slightly
above the G-band share it brackets, so the catalog arm is a lower bound on the true
grid share.

The samples that anchor the reliability audit are small and footprint-limited:
159 spectroscopic confirmations, 127 of them with GALEX coverage, and an
off-sequence UV-deficiency that rests on the 383 off-sequence sources with a GALEX
AIS NUV match within $3''$ (FUV detection is then the measured outcome). Those 383
are 10.4\% of the NUV-matched sample while off-sequence sources are 22.8\% of the
catalog, so the NUV match reaches them at less than half the rate it reaches the
rest, and the control behind the 0.57 matches distance only, not apparent
magnitude. GALEX covers 12.2\%
of the catalog with a UV-bright, nearby bias, the magnitudes are observed and
carry no dereddening, and the FUV is at AIS depth, so the UV indicator is
demonstrated on that subset and reads absence of a hot white dwarf. The 47\% purity floor is
for the NUV-bright GALEX-AIS-matched subset only. The spectroscopic samples that
anchor Table~\ref{tab:external} carry their own selection function, which favors
systems with an optically visible white dwarf, so their reach is not a random
sample of the catalog either.

\section{Conclusion}
\label{sec:conclusion}

\begin{itemize}
\item Held to the BP-band residual that survives XP correction, better than 2\%
below 400~nm and better than 1\% redward, with our blue taper clipped to that
bound, the
injection spurious rate is $0.09\pm0.02$ against a 0.053 baseline under the
templates' published per-pixel flux errors, elevated in 19 of 20
seeds: about 1.7 times the baseline. Letting the taper
hold 2\% past the point where the bound drops to 1\% gives $0.11\pm0.03$, and a
flat SNR-30 model on that same taper gives $0.08\pm0.01$. On the flat-model leg
the amortized posterior's 90\% coverage is 0.85, against 0.88 clean, though
the PIT rank test rejects uniformity there. Both
degrade far above the corrected residual, to a 0.96 spurious rate and 0.33
coverage near a 50\% local excess. On
these templates, 2\% of total flux put into the blue band comes out near a factor
of 27 above the same 2\% read per wavelength.

\item Bulk manufacture needs a local excess of 10--15\%
(Figure~\ref{fig:turnon}). Outside the color--magnitude box Huang state their
correction for, 40\% of the catalog, we have no measurement of the residual at
all, so the result there is a caution and nothing firmer. Where the systematic is
large enough to make spurious candidates, the plain goodness-of-fit partly flags
them while the $\Delta\chi^2$ prefers them (AUC 0.90 at share 0.01, near the
catalog's median of 0.0096, and 0.61 when shares follow the catalog's own
distribution, both at a 20\% local excess; 0.71 and 0.51 scoring all the injected
singles). The model-comparison gate protects the selection only for
binaries with a WD flux share above about 0.05. At share 0.01 it sits close to
chance, 0.43 conditioned and 0.54 unconditioned, and on the catalog's own share
mix it is weakly informative at best, 0.58 against 0.66. Either way it rejects
most genuine binaries, so its protection does not reach the catalog's typical
candidate.

\item When a blue calibration residual does miscalibrate an amortized posterior,
the failure is spectrally specific: it drives a tight but biased companion
fraction, whose 90\% coverage falls to 0.04 under a blue residual of 0.1124 local
across four seeds, against 0.27 under a grey offset, 0.54 under a red excess, and
0.67 under extra noise, each tuned to the same coverage damage. That blue
amplitude is an 11\% local excess, 5.6 times the residual audited above, so these
four arms describe what a large residual does to the posterior. They say nothing
about it at Huang's bound. The fit quality elsewhere stays clean, and a
posterior-predictive check on the blackbody proxy retains only partial power
(AUC 0.79, one seed, at about 23\% local on that population). This does not bear
on Li et al.'s Gaussian-process probability.

\item Independently of the injection, the catalog's external coverage is below
1\%, the prior-driven majority ($\sim$56\% with $\Delta\chi^2<0$) is externally
unverified, and the one clean, UV-validated contamination signal is
off-cooling-sequence sources (FUV-detected at 0.19 against 0.50, and it survives
a distance control). The probability appears to rank white-dwarf presence, on an
uncontrolled tercile comparison, without certifying
it, and does not control FDR at $q\le0.2$.
\end{itemize}

\section*{Data and code availability}
The audited catalog is the FaintQC sample of \citet{li2025wdms}, released on
Zenodo at \url{https://doi.org/10.5281/zenodo.14569070}. That is the version DOI, and
the concept DOI \url{https://doi.org/10.5281/zenodo.14411002} resolves to the latest
version. The analysis code and
configurations are at \url{https://github.com/WizardEternal/wdms-xp-calibration},
and every number in this paper reads off a committed record, with the three
exceptions named in Section~\ref{sec:limitations}. The 240 XP template spectra are not redistributed here; they are
re-downloadable from the Gaia archive by \texttt{source\_id} with the committed
fetch script and manifest.

\section*{Acknowledgments}
I thank Vaibhav Yenare for introducing me to white-dwarf main-sequence
binaries and the Gaia XP literature on them, including the Li et al.\ catalog
audited here.
This work used Gaia DR3 (ESA), the SDSS and LAMOST spectroscopic archives, and
GALEX. The analysis used numpy, scipy, matplotlib, astroquery, sbi
\citep{tejero-cantero2020sbi}, and UltraNest \citep{buchner2021ultranest}.

\appendix

\section{Methods detail}
\label{app:methods}

This appendix gives the exact forward model, fit library, gate statistics, and
calibration diagnostics, at the level needed to reproduce the results without the
code. Every spectrum is resampled onto one working grid, 61 pixels at
$\lambda = 392, 402, \dots, 992$~nm with 10~nm spacing, which is narrower than
the 336--1020~nm span the released \texttt{XP\_SAMPLED} products carry. Throughout,
each template and each synthetic spectrum is normalized to unit mean flux over
this grid before fitting, so amplitudes are dimensionless.

\subsection{The BP-band injection}
\label{app:inject}

The audited systematic is the blue-flux residual the corrected XP spectra keep
below 400~nm (\citealt{montegriffo2023xp}; \citealt{huang2024xp}). The related
$G_{\rm BP}$ photometric bias for faint red sources is a different effect, and it
is the one Li et al.'s $B<18$ selection targets (\citealt{riello2021gaia};
\citealt{li2025wdms}, their Sec.~2.2). A real white-dwarf companion adds
a smooth hot continuum across the whole optical (both BP and RP); the BP
systematic adds flux only in the blue. We model it as a half-cosine taper that is
1 at the blue edge of the BP band, falls to 0 at 680~nm, and is identically zero in
RP. The shape peaks at the 392~nm blue edge and dies by 680~nm, tracking the
wavelength dependence of the residual \citet{huang2024xp} report, which is worst
shortward of 400~nm. Writing $x \equiv \mathrm{clip}\!\left[(\lambda - 392)/(680 - 392),\,0,\,1\right]$
with $\lambda$ in nm, the unnormalized shape is
\begin{equation}
\tilde r(\lambda) =
\begin{cases}
\cos^2\!\left(\tfrac{\pi}{2}\,x\right), & \lambda \le 680~\mathrm{nm},\\[4pt]
0, & \lambda > 680~\mathrm{nm},
\end{cases}
\end{equation}
and the injection template is normalized to unit mean over the 61-pixel grid,
\begin{equation}
\mathrm{RINJ}(\lambda) = \tilde r(\lambda) \,/\, \overline{\tilde r},
\qquad
\overline{\tilde r} = \frac{1}{N_\lambda}\sum_{\lambda} \tilde r(\lambda),
\qquad N_\lambda = 61 .
\end{equation}

\paragraph{Severity to local excess.}
\label{app:severity}
A clean single-star spectrum $x_s$ can be contaminated by adding the template
scaled to a fraction of the spectrum's own \emph{mean} flux. With dimensionless
severity $s$,
\begin{equation}
x_t = x_s + s\,\langle x_s\rangle\,\mathrm{RINJ},
\qquad
\langle x_s\rangle = \frac{1}{N_\lambda}\sum_\lambda x_s(\lambda) .
\label{eq:severity}
\end{equation}
Because RINJ has unit mean, the injected flux integrated over the grid is exactly a
fraction $s$ of the spectrum's total flux, all of it deposited blueward of 680~nm.
That is not the local, per-wavelength residual \citet{huang2024xp} report. RINJ
peaks at about 4 times its mean at 392~nm, and the blue flux of a red MS star is a
small part of its total, so $s = 0.02$ of the total is a median 55\% \emph{local}
excess in the blue across the real MS templates (the per-template median over
392--420~nm, taken across the 120 templates; 78\% at the 392~nm edge), about 27
times a 2\% local residual. Our resampled grid starts at 392~nm, so of these pixels
only 392~nm lies inside the 336--400~nm range Huang quote their 2\% for, and that
78\% edge value is the in-band figure, 39 times a 2\% local residual. Pooling the
pixels instead of the templates gives a median of about 59\%, or 30 times. The
three aggregations run 27, 30 and 39 times and we quote the smallest of them
throughout. The conversion is a property of the spectra it is measured on, so it
covers the real-template runs only; the two runs that inject into a blackbody
proxy carry their own, given in Appendix~\ref{app:evidence} and
Section~\ref{sec:discussion}.
The turn-on of Section~\ref{sec:silent} and the
coverage curve are therefore run with a multiplicative-local injection,
$x_t = x_s\,(1 + a\,\hat r)$, where $\hat r$ is the 0-to-1 blue taper and $a$ is the
local fractional excess, so the plotted amplitude is the local excess. The
additive form of Eq.~\eqref{eq:severity} is kept only
to describe the large-amplitude regime.

\paragraph{The clipped profile.}
\citet{huang2024xp} quote their corrected-XP consistency in two parts, better
than 2\% over 336--400~nm and better than 1\% redward, so a taper whose amplitude
is set from the 2\% half alone breaches the 1\% half over part of its span. The
turn-on of Section~\ref{sec:silent} therefore uses the clipped local excess
\begin{equation}
e(\lambda) =
\begin{cases}
a\,\hat r(\lambda), & \lambda \le 400~\mathrm{nm},\\[3pt]
\min\!\left[a\,\hat r(\lambda),\, 0.01\right], & \lambda > 400~\mathrm{nm},
\end{cases}
\qquad x_t = x_s\,\bigl(1 + e\bigr),
\label{eq:huangenv}
\end{equation}
which sits on the bound wherever the taper would exceed it and below it
elsewhere. At $a = 0.02$ that is 2\% at 392~nm and exactly 1\% from 402 to
532~nm, falling with the taper beyond. That is the largest version of this taper
the bound allows. Other shapes under the same bound go further: the clipped
profile carries nothing past 680~nm, where the 1\% cap still applies, so a shape
that held the cap across the whole grid would inject more. We do not run
one. The unclipped taper at the same $a$ is
kept as the upper-bound arm, and a third cell runs the unclipped taper at
$a = 0.01$, the largest amplitude at which it stays inside both halves of the
bound at every pixel.

\paragraph{Noise models.}
Two noise models are used, both independent Gaussian per pixel. The first has a
single spectrum-level scale. For a
noise-free spectrum $x_0$ at signal-to-noise SNR (default 30),
\begin{equation}
\sigma = \langle x_0\rangle / \mathrm{SNR},
\qquad
x = x_0 + \sigma\,\epsilon, \quad \epsilon \sim \mathcal{N}(0, \mathbb{1}_{N_\lambda}) ,
\end{equation}
so $\sigma$ is constant across the grid for a given object and equal to its mean
flux over the SNR. The same $\sigma$ enters the $\chi^2$ in all fits. For the
injection runs the noise is drawn on the clean spectrum and the excess, in
either form above, is then applied to the noisy spectrum, with $\sigma$ kept at
its clean value, so the noise realization is shared between the clean and
contaminated copies of each object (the multiplicative form also scales the
noise in the pixels it touches).

The turn-on of Section~\ref{sec:silent} uses the second model, built from the
published errors: for each template, the fractional error, the ratio of
\texttt{flux\_error} to \texttt{flux},
is interpolated onto the 61-pixel grid and multiplied by the
model flux, with no reduction for resampling (the XP samples are correlated, so
an averaging credit would overclaim independence). The resulting $\sigma_i$
vector enters noise generation, the NNLS weighting, $\Delta\chi^2$, and the
95th-percentile threshold, which recalibrates automatically. The flat
$\sigma=\langle x_0\rangle/\mathrm{SNR}$ model above is retained as a
model-dependence arm; the PCA leg of Appendix~\ref{app:pca} uses the flat model
throughout.

\subsection{The fit library and the model-comparison gate}
\label{app:gate}

\paragraph{Library.}
Templates are read from disk (real single MS and single WD Gaia XP spectra, or
theoretical grids), linearly interpolated onto the 61-pixel grid, dropped if they
do not cover the full grid or have nonpositive mean, and mean-normalized. The
binary fit is $O(N_{\rm MS}\,N_{\rm WD})$, so when a class has more than 50 members
the fit library is capped at a representative random subsample of 50 (the larger
real-spectra libraries hold 120 per class; the 50-cap applies to the fitting step,
the PCA model of Appendix~\ref{app:pca} uses all 120). The combined library is
$\mathrm{ALL} = [\mathrm{MS}; \mathrm{WD}]$ with MS templates first. The scored
objects come from the same subsample: the $n = 150$ clean singles of a turn-on
seed are drawn from its 50 MS templates with replacement, so a template can be
scored several times in one seed and fewer than 50 distinct spectra stand behind
the 150.

\paragraph{Fits.}
All amplitudes are non-negative, solved by non-negative least squares (NNLS) on the
noise-weighted design matrix, i.e. each basis row and the data are divided by
$\sigma$ before the solve. Three fits are computed for every spectrum:
\begin{itemize}
\item \emph{Single} ($k=2$ free parameters: template index plus one amplitude):
the best single template over ALL,
$\chi^2_{\rm s} = \min_t \chi^2\!\left(a\,T_t\right)$.
\item \emph{Binary} ($k=4$: two template indices plus two amplitudes): the best
MS$+$WD pair,
$\chi^2_{\rm b} = \min_{i,j}\chi^2\!\left(c_1\,\mathrm{MS}_i + c_2\,\mathrm{WD}_j\right)$.
\item \emph{Single$+$systematic} ($k=3$: MS template index plus two amplitudes):
the best MS template plus a fixed blue-artifact basis,
$\chi^2_{\rm r} = \min_i \chi^2\!\left(c_1\,\mathrm{MS}_i + c_2\,B\right)$.
\end{itemize}
Here $\chi^2(m) = \sum_\lambda \big[(x(\lambda) - m(\lambda))/\sigma\big]^2$.

\paragraph{Leave-one-out.}
To avoid trivially perfect fits, each synthetic truth spectrum is fit with its own
generating template(s) excluded from the library: a single drawn from MS template
$i$ excludes index $i$; a binary drawn from $(i,j)$ excludes both. The
severity-sweep run reports each metric both without leave-one-out and with it;
the share-resolved runs of Section~\ref{sec:silent} are leave-one-out
throughout. All quoted fit-library values are leave-one-out.

\paragraph{Renormalized $\Delta\chi^2$ and the binary threshold.}
The per-object single-versus-binary statistic is the fractional $\chi^2$
improvement of the binary fit,
\begin{equation}
d = \frac{\chi^2_{\rm s} - \chi^2_{\rm b}}{\chi^2_{\rm s}} \in [0,1) .
\label{eq:dchi2}
\end{equation}
This renormalization removes the per-object noise scale so the statistic is
comparable across spectra. The statistic $d$ is not Li et al.'s
\texttt{chi2\_diff\_renorm} column, which is a different renormalization of the
same comparison and runs negative for 56\% of their rows. The binary-decision
threshold is the 95th percentile of
$d$ over the clean single-star population, $\theta = Q_{0.95}(d_{\rm s})$, i.e. a
5\% false-positive rate on clean singles by construction. A spectrum is flagged as a
candidate binary when $d > \theta$; the spurious rate is the fraction of
BP-injected singles that cross $\theta$. With $n = 150$ clean singles this lands
on 8 of them in every seed, so the baseline is the constant 0.053 and the
threshold carries all the seed-to-seed variation (0.29--0.55). The turn-on runs
take that percentile in sample, on the same clean singles whose injected copies
are then scored, which is what makes the severity-zero point exactly 0.053; the
share-resolved gate and AUC runs of Section~\ref{sec:silent} instead set it on an
independent clean control drawn without injection.

\paragraph{The systematic-aware gate.}
The model-comparison gate replaces the per-object $\Delta\chi^2$ with a BIC
difference between the binary model and the single$+$systematic model. With
$N_\lambda = 61$ pixels and $\mathrm{BIC}(\chi^2, k) = \chi^2 + k\ln N_\lambda$,
the gate statistic is
\begin{equation}
G = \mathrm{BIC}(\chi^2_{\rm b}, 4) - \mathrm{BIC}(\chi^2_{\rm r}, 3)
  = \chi^2_{\rm b} - \chi^2_{\rm r} + \ln N_\lambda .
\label{eq:gate}
\end{equation}
A large $G$ means the binary model is not preferred once the blue artifact is
allowed for, i.e. the object reads as a systematic. The
$k=4$ versus $k=3$ parameter counts give the constant $+\ln N_\lambda \approx 4.11$
offset. The gate's artifact basis $B$ is not the injected shape.
To avoid an oracle confound the gate uses a different but plausible blue ramp,
\begin{equation}
\tilde g(\lambda) =
\begin{cases}
\mathrm{clip}\!\left[(680 - \lambda)/(680 - 392),\,0,\,1\right], & \lambda \le 680~\mathrm{nm},\\
0, & \lambda > 680~\mathrm{nm},
\end{cases}
\qquad
\mathrm{RGATE} = \tilde g / \overline{\tilde g} ,
\end{equation}
a linear taper (1 at 392~nm, 0 at 680~nm), mean-normalized, versus the half-cosine
RINJ that is actually injected. Every gate number quoted here is the mismatched
one. The committed records do carry a matched-basis oracle variant, unused here.

\subsection{The discrete-rank SBC null}
\label{app:sbc}

We assess calibration with simulation-based calibration (SBC; \citealt{talts2018sbc}).
For each of $N_{\rm SBC} = 300$ test objects we draw a parameter vector
$\theta^\star$ from the prior, simulate an observation, and draw $L = 50$ samples
from the amortized posterior. The rank of $\theta^\star$ in those draws is computed
componentwise as the number of posterior samples below the truth,
\begin{equation}
\mathrm{rank}_j = \#\{\,\theta^{(\ell)}_j < \theta^\star_j\,\} \in \{0, 1, \dots, L\} ,
\end{equation}
and the probability integral transform is $\mathrm{PIT}_j = \mathrm{rank}_j / L$.

Posterior draws are taken without rejection, with
\texttt{reject\_outside\_prior} set to \texttt{False}: the default rejection sampler hangs on the
out-of-distribution contaminated inputs, and to \emph{measure} miscalibration we
want the flow's actual draws, not draws forced back inside the prior. If the
posterior is calibrated, the ranks are discrete-uniform on $\{0, \dots, L\}$.

\paragraph{Coverage and its discrete null.}
We summarize each parameter by its central-90\% coverage, the fraction of test
objects with $0.05 \le \mathrm{PIT}_j \le 0.95$, and report the mean over
parameters. With $L = 50$ samples the rank takes $L + 1 = 51$ equally likely
integer values. The PIT band $[0.05, 0.95]$ maps to rank values in
$[2.5, 47.5]$, i.e. the integer ranks $3, 4, \dots, 47$, which is
$47 - 3 + 1 = 45$ of the 51 values. So the expected coverage under perfect
calibration is not 0.90 but the discrete-rank null
\begin{equation}
\mathrm{cov}_{90}^{\rm null} = \frac{45}{51} = 0.8824 ,
\label{eq:covnull}
\end{equation}
which is the reference line our clean-data coverage is compared against; a coverage
well below this, together with a KS test of PIT-uniformity driven to small
$p$, signals miscalibration. We also report the per-parameter Kolmogorov-Smirnov
$p$-value of $\mathrm{PIT}_j$ against the uniform distribution and its minimum over
parameters. On clean data that minimum reads 0.101, 0.041, 0.015 and 0.075 over
the four seeds, which is what a minimum over six parameters does under perfect
calibration: $P(\min < 0.05) = 1 - 0.95^{6} = 0.26$, so two of four below 0.05 is
unremarkable. Under the injection at $a = 0.02$ it falls below $10^{-5}$ in all
four seeds, while the mean coverage over the same step moves only from 0.879 to
0.854. The coverage is what the turn-on curve plots, so that is the number we
quote, and the KS statistic is the sharper of the two.
The diagnostic is run on clean held-out simulations and, separately, on
simulations with the BP systematic injected at inference time as the multiplicative-local
excess of Appendix~\ref{app:severity} (local amplitude $a$); the
training set never contains the systematic.

\subsection{The real-template PCA forward model}
\label{app:pca}

For the SBC diagnostic we need a smooth, differentiable map from parameters to
spectra. The blackbody-proxy model (MS blackbody $+$ WD blackbody projected onto
the grid, with the four parameters $T_{\rm ms}, \log_{10}a_{\rm ms}, T_{\rm wd},
\log_{10}a_{\rm wd}$) is the simplest such map. To ground the diagnostic in real
spectra we also build a PCA model directly from the Gaia XP template libraries
(120 single MS and 120 single WD), as follows.

\paragraph{PCA construction.}
For each class we mean-center the library $L$ and take the singular value
decomposition $L - \mu = U\,S\,V^\top$. The first $k$ right-singular vectors are the
principal components $\mathrm{PC} = V^\top_{1:k}$ and the scores are
$S = (L - \mu)\,\mathrm{PC}^\top$. We keep $k_{\rm MS} = 2$ components for the MS
library (more than 99\% of the variance) and $k_{\rm WD} = 3$ for the WD library
(about 96\% of the variance). A spectrum of each class is then reconstructed as
\begin{equation}
\mathrm{MS}(s_{\rm ms}) = \mu_{\rm ms} + s_{\rm ms}\cdot\mathrm{PC}_{\rm ms},
\qquad
\mathrm{WD}(s_{\rm wd}) = \mu_{\rm wd} + s_{\rm wd}\cdot\mathrm{PC}_{\rm wd} .
\end{equation}

\paragraph{Parametrization and prior.}
The six free parameters are the PC scores plus a log mixing fraction,
\begin{equation}
\theta = (s_{\rm ms,1},\, s_{\rm ms,2},\, s_{\rm wd,1},\, s_{\rm wd,2},\, s_{\rm wd,3},\, \log_{10}f) .
\end{equation}
The prior is a box. Each score component ranges over the empirical
$[\min, \max]$ spanned by the real templates. The box strictly contains the
template hull, so a prior draw can sit off the template manifold while staying
inside the observed per-component ranges; reconstructions are strictly positive
over the whole box (closed-form corner minima 0.038 for MS and 0.085 for WD, in
unit-mean flux). The mixing fraction ranges over
$\log_{10}f \in [\log_{10}0.05,\, \log_{10}0.30]$, matching the WD flux fraction of
the proxy model.

\paragraph{Combination and injection.}
The white dwarf is flux-matched to the MS total flux and scaled by $f$ before
addition, then noised at SNR 30,
\begin{equation}
x_0 = \mathrm{MS}(\theta) + f\,\frac{\sum_\lambda \mathrm{MS}(\theta)}{\sum_\lambda \mathrm{WD}(\theta)}\,\mathrm{WD}(\theta),
\qquad f = 10^{\log_{10}f},
\end{equation}
with $\sigma = \langle x_0\rangle / \mathrm{SNR}$ and Gaussian per-pixel noise as in
Appendix~\ref{app:inject}. At inference the BP systematic is injected as the
multiplicative-local excess of Appendix~\ref{app:severity} with local amplitude
$a$: the noise-free spectrum is scaled, $x_0 \to x_0\,(1 + a\,\hat r)$, the noise
level is recomputed from the contaminated mean, and the noise is drawn on top.
The systematic is absent from training, and the reported coverage curve sweeps
$a$ over 2--50\% local. This chain differs from the injection leg of
Section~\ref{sec:silent}, which applies the excess to the already-noisy spectrum
at clean $\sigma$; the two legs of Figure~\ref{fig:turnon} are matched in
amplitude, not in noise chain. The PCA leg also has no leave-one-out analogue:
reconstruction uses the full 120-template basis, while the fit-library leg
excludes generating templates. The amortized posterior is a neural spline
flow (sbi NPE), trained on $N_{\rm train} = 8000$ prior draws and their simulated
spectra; SBC then proceeds as in Appendix~\ref{app:sbc}.

\subsection{The nested-sampling evidence gate}
\label{app:evidence}

To confirm the BIC-proxy gate against a proper marginal likelihood we run a small
nested-sampling experiment (UltraNest) on the blackbody proxy, $n = 15$ objects per
class. For each spectrum we compute the log-evidence $\ln Z$ of two models with flat
priors on the stated ranges:
\begin{itemize}
\item \emph{Binary} (4 parameters): $m = 10^{\log a_{\rm ms}}\mathrm{BB}(T_{\rm ms}) + 10^{\log a_{\rm wd}}\mathrm{BB}(T_{\rm wd})$,
with $T_{\rm ms} \in [3000, 7000]$~K, $T_{\rm wd} \in [7000, 40000]$~K,
$\log a_{\rm ms} \in [-1, 1]$, $\log a_{\rm wd} \in [-3, 1]$.
\item \emph{Single$+$systematic} (3 parameters): $m = 10^{\log a}\mathrm{BB}(T) + 10^{\log b}\,\mathrm{RGATE}$,
with $T \in [3000, 7000]$~K, $\log a \in [-1, 1]$, $\log b \in [-3, 1]$, using the
mismatched RGATE basis.
\end{itemize}
The Gaussian likelihood is $\ln\mathcal{L} = -\tfrac12\sum_\lambda
\big[(x - m)/\sigma\big]^2$ and each run uses 60 live points. The gate statistic is
$\Delta\ln Z = \ln Z_{\rm binary} - \ln Z_{\rm single+sys}$.

The two committed runs use the additive severities $s = 0.02$ and $s = 0.05$ of
Eq.~\eqref{eq:severity}. Measured on the Planck single-star proxy these runs
actually inject into, those are local excesses of about 11\% and 28\%, so
both sit well above anything the correction leaves,
and the genuine binaries are drawn in the bright band (WD flux ratio
0.05--0.30). At $s = 0.02$ the median $\Delta\ln Z$ is
$+15.4$ for genuine binaries and $-0.20$ for BP-injected singles, so the sign
separates the two classes while the spurious side sits close to indifference.
Ranking the spurious against the genuine by $-\Delta\ln Z$ gives AUC 0.94, against
0.12 when the same labels are ranked by the per-object $\Delta\chi^2$ of
Eq.~\eqref{eq:dchi2}, which is blind to the systematic. The gate runs on the
blackbody proxy, not the real templates, so it confirms only the sign of the BIC
result.

\bibliographystyle{plainnat}
\bibliography{note}

\end{document}